\let\oldvec\vec
\documentclass[runningheads]{llncs}
\let\vec\oldvec

\usepackage{times}  
\usepackage{helvet}  
\usepackage{courier}  
\usepackage{graphicx}  
\usepackage{amssymb}
\usepackage{amsmath}
\usepackage[group-separator={,}]{siunitx}
\usepackage{multirow}
\usepackage{tikz}

\newtheorem{mytheorem}{Theorem}
\newtheorem{mydefinition}{Definition}

\newtheorem{myexample}{Example}

\newcommand{\myqed}{\mbox{$\diamond$}}
\newcommand{\myimp}{\mbox{$\Join$}}
\newcolumntype{C}{>{\centering\arraybackslash}p{1.5cm}}

\newcommand{\mylike}{\textsc{Li\-ke}}
\newcommand{\myblike}{\textsc{Ba\-la\-n\-ced Li\-ke}}
\newcommand{\mymaxlike}{\textsc{Ma\-xi\-mum Li\-ke}}
\newcommand{\myminlike}{\textsc{Mi\-ni\-mum Li\-ke}}
\newcommand{\myminutil}{\textsc{Mi\-ni\-mum Uti\-li\-ty}}

\begin{document}

\title{Monotone and Online Fair Division}
\titlerunning{Monotone and Online Fair Division}

\author{Martin Aleksandrov \and Toby Walsh}
\authorrunning{M. Aleksandrov \and T. Walsh}

\institute{Technical University Berlin, Berlin, Germany \\ \{martin.aleksandrov,toby.walsh\}@tu-berlin.de}

\maketitle

\begin{abstract}
We study a new but simple model for online fair division in which indivisible items arrive one-by-one and agents have monotone utilities over bundles of the items. We consider axiomatic properties of mechanisms for this model such as strategy-proofness, envy-freeness and Pareto efficiency. We prove a number of impossibility results that justify why we consider relaxations of the properties, as well as why we consider restricted preference domains on which good axiomatic properties can be achieved. We propose two mechanisms that have good axiomatic fairness properties on restricted but common preference domains.
\end{abstract}

\section{Introduction}\label{sec:intro}

Many studies of fair division problems make some simplifying assumptions such as: the problem is offline (i.e.\ the items and agents are all simultaneously available), and agents have additive utilities over the items. In practice, however, such assumptions may be violated. Recently, Walsh \cite{walsh2014} introduced a simple {\em online} model for the fair divison of indivisible items in which, whilst utilities remain additive, the items become available over time and must be allocated to agents immediately. Such an online model has many practical applications. For example, donated kidneys must be allocated to  patients as they become available. As a second example, places on university courses open each term and must be allocated before classes begin, and before places open for the following term. As a third example, a charging station might be allocated to a waiting electric car immediately it is freed up. And, as a fourth example, perishable items donated to a food bank might have to be allocated to charities feeding the poor immediately. As a fifth example, when allocating memory to cloud services, we may not know what and how many services are requested in the next moment.

In this paper, we relax this model of online fair division to deal with monotone utilities. There are many settings where utilities might not be additive. For instance, agents may have diminishing returns for multiple copies of an item. You may, for example, gain less utility for a second bicycle. Agents may also have complementarities. You may, for example, get little utility for the cricket bat unless you also get  the cricket ball. We thus consider a model of online fair division in which agents have monotone but possibly non-additive utilities. Indeed, monotone utilities are especially challenging in an online setting. As utilities may not be additive, we cannot allocate items independently of previous or, more problematically, of future items. Suppose agent 1 only likes item $a$ in the presence of item $b$, whilst agent 2 only likes $a$ in the presence of $c$. Then the decision to give item $a$ to agent 1 or 2 may depend on whether items $b$ or $c$ will arrive in the future, which we suppose is not known.

We define firstly the model of online fair division with monotone utilities and propose to consider non-wasteful marginal mechanisms for it. We then show that no non-wasteful mechanism can guarantee simple axiomatic properties such as strategy-proofness, envy-freeness (even approximately) or Pareto efficiency under weak conditions, whilst that was possible with additive utilities. We then consider monotone utilities with non-zero marginals. In the offline setting, this is a natural class of restricted preferences in which agents are assumed to prefer always having an item to not having it, supposing that their marginal utility for it could be arbitrarily small. We prove that many axiomatic properties can be achieved in this domain. We also consider a weaker form of strategy-proofness adapted to our online setting that supposes agents only have knowledge of the current item, and not of any future items that might or might not arrive. Finally, we propose two mechanisms - the \myminlike\ and \myminutil\ mechanisms - and prove that they satisfy this weaker form of strategy-proofness as well as envy-freeness up to some item in common domains with identical utilities. 

\section{Related Work}\label{sec:related}

Our model of online fair division with monotone utilities
generalizes an existing model of online fair division with additive utilities
introduced in \cite{walsh2014}. Aleksandrov et al. \cite{aleksandrov2015ijcai} analysed two simple randomized mechanisms for this model, called \mylike\ and \myblike. The \mylike\ mechanism allocates an incoming item uniformly at random to one of the agents that declares non-zero bid for it. This is strategy-proof and envy-free in expectation. The \myblike\ mechanism allocates an incoming item to an agent with the fewest items currently amongst those that declare non-zero bids for the incoming item. With 0/1 utilities, this bounds the envy of agents, and is strategy-proof for 2 but not more agents. Some other online mechanisms (e.g.\ \mymaxlike) that are Pareto efficient ex post and ex ante are considered in \cite{aleksandrov2017mcm}. We can extend these to mechanisms for monotone but not necessarily additive utilities by allocating an incoming item to one of the agents that declares a non-zero marginal bid for the item. However, we prove that none of these mechanisms or even any other mechanism is strategy-proof, envy-free or Pareto efficient in our setting with monotone utilities. 

Further, for the model with additive utilities, Benade et al. \cite{benade2018} showed that the random assignment of each next item (i.e.\ \mylike) diminishes the envy over time. By comparison, we prove that approximations of envy-freeness ex post such as EF1 (see \cite{budish2012}) and EFX (see \cite{caragiannis2016}) cannot be satisfied in our monotone setting. On the other hand, we further prove that EF1 can only be satisfied in two restricted but common preference domains of identical utilities. We also contrast our results with similar results in (offline) fair division. For example, it remains an open question if offline EFX allocations exist in general. We prove that no mechanism for online fair division can return such allocations even when they exist. This holds with identical additive utilities in which domain there are offline algorithms that return such allocations \cite{barman2018}. Further, we can show that some other (offline) characterizations (e.g.\ \cite{brams2005,manea2007}) break in the online setting. In contrast, our results can be mapped to offline settings as online mechanisms can be applied to offline problems by presenting the items in some (perhaps random) order.

There are other related models. For example, Walsh \cite{walsh2011} has proposed a different online model in which items are divisible (not indivisible) and agents (not items) arrive over time. Also, Kash, Procaccia and Shah \cite{kash2014} have proposed a dynamic model of fair division in which agents again arrive over time, but there are multiple homogeneous (not heterogeneous) and divisible items. There is also a connection between our consideration of marginal bidding and the one for auctions that has been made by Greenwald and Boyan in \cite{greenwald2004}. One interesting difference between our work and theirs is that marginal utility bidding is an optimal strategy for sequential auctions whereas, as we prove, it may not be for online mechanisms. Finally, other related works in fair division (e.g.\ \cite{aleksandrov2017eom,aleksandrov2017ijcai,freeman2018,hosseini2015a}), voting (e.g.\ \cite{chevaleyre2012,gibbard1973,xia2010}) and kidney exchange (e.g.\ \cite{dickerson2012,dickerson2013}) exist. However, our results do not follow from prior results.

\section{Monotone and Online Fair Division}\label{sec:model}

We consider an online fair division problem with \emph{agents} from $[n]=\lbrace 1,\ldots,n\rbrace$ and \emph{indivisible items} from $O=\lbrace o_1,\ldots,o_m\rbrace$, where $m\in\mathbb{N}_{\geq 1}$. WLOG, we suppose that items arrive one-by-one from $o_1$ to $o_m$. Thus, we write $O_j$ for the subset of $O$ of the first $j$ items. We suppose that agents have bundle utilities. We write $u_i(B)\in\mathbb{R}_{\geq 0}$ for the \emph{(private) utility} of $i\in [n]$ for each $B\subseteq O$. We also write $u_i(o)$ for $u_i(\lbrace o\rbrace)$. We suppose $u_i(\emptyset)=0$. We say that the agents have \emph{identical utilities} iff, for each $i,k\in [n]$ and $B\subseteq O$, $u_i(B)=u_k(B)$. In this case, we write $u(B)$ for $u_i(B)$. We further write $u_i(B\cup\lbrace o\rbrace)-u_i(B)$ for the \emph{marginal utility} of $i\in [n]$ for each $B\subset O$ and $o\in O\setminus B$. We say that this marginal utility is \emph{general} iff $u_i(B\cup\lbrace o\rbrace)-u_i(B)\in\mathbb{R}_{\geq 0}$, and \emph{non-zero} iff $u_i(B\cup\lbrace o\rbrace)-u_i(B)\in\mathbb{R}_{>0}$. We write $\pi=(\pi_{1},\ldots,\pi_{n})$ for an allocation of the items from $B$ to the agents, where $\cup_{i\in [n]} \pi_i = B$ and $\pi_{i}\cap \pi_{j}=\emptyset$ for $i,j\in [n]$ with $i\neq j$. And, we let $\Pi_j=\lbrace \pi| \pi\mbox{ is an allocation with }\cup_{i\in [n]} \pi_i = O_j\rbrace$. 

We consider \emph{online} mechanisms that allocate each next item without the knowledge of any future items. We focus on \emph{non-wasteful} mechanisms that allocate items to agents that declare non-zero marginal utility for item $o_j$, if there are such agents. At round 1, each agent $i\in [n]$ becomes aware of their marginal utility $u_i(o_1)$ for $o_1$. And, at round $j$th ($j>1$), each agent $i$ becomes aware of their marginal utility $u_i=u_i(\pi_i\cup\lbrace o_j\rbrace)-u_i(\pi_i)$ for $o_j$ where $\pi\in \Pi_{j-1}$ is some allocation of the first $(j-1)$ items. The mechanism firstly asks each $i\in [n]$ for a marginal bid $v_i$ for $o_j$. Agents may act strategically and bid insincerely, i.e.\ $v_i$ may be different from $u_i$. We say that $i$ \emph{likes} $o_j$ if $u_i>0$. The mechanism secondly shares the probability of 1 for $o_j$ among those who make non-zero marginal bids. If there are no such agents, $o_j$ is allocated at random.

A mechanism thus returns a probability distribution over the allocations in $\Pi_j$. We write $\Delta_j=(p(\pi)|\pi\in\Pi_j)$ for it, where $p(\pi)\in [0,1]$ is the probability of $\pi\in\Pi_{j}$. We have that $\sum_{\pi\in\Pi_j}p(\pi)=1$. We write $u_{i}(\pi_k)$ for the \emph{monotone utility} of agent $i$ for the items of agent $k$ in $\pi$. We write $\overline{u}_{ik}(\Pi_{j})$ for the \emph{expected utility} of agent $i$ for the expected allocation of agent $k$ in $\Pi_j$. We have $\overline{u}_{ik}(\Pi_{j})=\sum_{\pi\in\Pi_j} p(\pi)\cdot u_i(\pi_k)$. We also write sometime $u_{i}(\pi)$ for $u_{i}(\pi_i)$ and $\overline{u}_{i}(\Pi_j)$ for $\overline{u}_{ii}(\Pi_j)$. Finally, we say that $u_i(\pi_k)$ is \emph{additive} iff it is $\sum_{o\in\pi_k} u_i(o)$. In this case, the expected utility of agent $i$ for the expected allocation of agent $k$ in $\Pi_j$ is also additive. That is, $\overline{u}_{ik}(\Pi_{j})=\overline{u}_{ik}(\Pi_{j-1})+\sum_{\pi\in\Pi_j} p(\pi)\cdot u_i(o_j)$.

\section{Axiomatic Properties}\label{sec:properties}

Three fundamental axiomatic properties of mechanisms for our setting concern the incentives of agents to bid strategically for an allocation, the fairness of an allocation and the economic efficiency of an allocation.

\begin{mydefinition} $($\emph{Strategy-proofness, SP}$)$
A mechanism is \emph{SP} in a problem with $m$ items if, with complete information of $o_1$ to $o_m$, no agent $i$ can strictly increase $\overline{u}_{i}(\Pi_m)$ by misreporting $u_i(\pi_i\cup\lbrace o_j\rbrace)-u_i(\pi_i)$ for one or more item $o_j$ and allocation $\pi\in\Pi_{j-1}$, supposing that every other agent $k\neq i$ bid sincerely their marginal utilities for items $o_1$ to $o_m$.
\end{mydefinition}

\begin{mydefinition}$($\emph{Envy-freeness, EF}$)$
A mechanism is \emph{EF ex post (EFP)} in a problem with $m$ items if, for each $\pi\in \Pi_m$ with $p(\pi)>0$, no agent $i$ envies another agent $k$, i.e.\ $\forall i,k : u_{i}(\pi_i)\geq u_{i}(\pi_k)$. It is {\em EF ex ante (EFA)} in a problem with $m$ items if no agent $i$ envies another agent $k$ in expectation, i.e.\ $\forall i,k : \overline{u}_{ii}(\Pi_m)\geq \overline{u}_{ik}(\Pi_m)$.
\end{mydefinition}

\begin{mydefinition}$($\emph{Pareto efficiency, PE}$)$
A mechanism is \emph{PE ex post (PEP)} in a problem with $m$ items if, for each $\pi\in \Pi_m$ with $p(\pi)>0$, no $\pi^{\prime}\in\Pi_m$ is such that $\forall i: u _{i}(\pi^{\prime}_i)\geq  u_{i}(\pi_i)$ and $\exists k:u_{k}(\pi^{\prime}_k)>u_{k}(\pi_k)$. It is \emph{PE ex ante (PEA)} in a problem with $m$ items if, no other probability distribution over the allocations in $\Pi_m$ gives to each agent $i$ at least $\overline{u}_{i}(\Pi_m)$ and to some agent $k$ strictly more than $\overline{u}_{k}(\Pi_m)$. 
\end{mydefinition}

We say that a mechanism satisfies a given property {\sc\bf P} iff, for each $m\in \mathbb{N}$, it satisfies {\sc\bf P} on each problem with $m$ items. We are interested in mechanisms for our model that satisfy combinations of these three properties. 

\section{General Marginal Utilities}\label{sec:imp}

We start with general marginal utilities. As we argued earlier, the monotone and online nature of our problem makes it more difficult to achieve nice axiomatic properties. Indeed, we will show that \emph{no} mechanism is strategy-proof, envy-free or Pareto efficient even in very limited utility domains, e.g.\ monotone utilities with binary marginals, identical monotone utilities, etc.

\subsection{Strategy-Proofness}

We prove firstly that strategy-proofness is impossible in general. The problem here is that the marginal utility of an agent for an item may depend on their allocation of past items, and thus so is their probability for the item (in a given allocation). We illustrate this in Example~\ref{exp:one}.

\begin{myexample}\label{exp:one}
Let us consider the online fair division problem with $2$ agents and $O=\lbrace o_1,o_2\rbrace$. Further, let $u_1(\emptyset)=0,u_1(\lbrace o_1\rbrace)=2,u_1(\lbrace o_2\rbrace)=4,u_1(O)=6$ and $u_2(\emptyset)=0,u_2(\lbrace o_1\rbrace)=5,u_2(\lbrace o_2\rbrace)=2,u_2(O)=5$. If agent 1 gets $o_1$, the marginal utilities of agents 1 and 2 for $o_2$ are $4$ (i.e.\ $u_1(O)-u_1(\lbrace o_1\rbrace)$) and $2$ (i.e.\ $u_2(\lbrace o_2\rbrace)-u_2(\emptyset)$). If agent 2 gets $o_1$, the marginal utilities of agents 1 and 2 for $o_2$ are $4$ (i.e.\ $u_1(\lbrace o_2\rbrace)-u_1(\emptyset)$) and $0$ (i.e.\ $u_2(O)-u_2(\lbrace o_1\rbrace)$). \myqed
\end{myexample}

It might, therefore, be beneficial for an agent to report strategically a marginal utility of  zero for the current item in order to increase their chance for their most favourite bundle of future items. Indeed, for this reason, \emph{no} mechanism is strategy-proof even with very restricted preferences. This contrasts with the case of additive utilities where, for example, the \mylike\ mechanism is strategy-proof \cite{aleksandrov2015ijcai}. 

\begin{mytheorem}\label{thm:one}
No non-wasteful mechanism for online fair division is strategy-proof, even with identical monotone utilities with 0/1 marginals.
\end{mytheorem}

\begin{myproof}
Consider agents 1 and 2, items $o_1$ to $o_3$ and ordering $(o_1,o_2,o_3)$. The utilities are identical for each $B\subseteq O$. If $|B|=1$, let $u(B)$ be $1$. If $|B|=2$, let $u(B)$ be $2$ if $B=\lbrace o_2,o_3\rbrace$ and $1$ otherwise. Also, let $u(O)=2$. Suppose agents are sincere and the mechanism gives $o_1$ to agent 1 with $p\in [0,1]$ and to agent 2 with $(1-p)$. We consider three cases. In the first case, the mechanism is randomized and $p\in (0,1)$. If it gives $o_1$ to agent 1 with $p$, then it gives $o_2$ and $o_3$ to agent 2 with probability $1$. If it gives $o_1$ to agent 2 with $(1-p)\in (0,1)$, then it gives $o_2$ and $o_3$ to agent 1 with probability $1$. Therefore, the expected utility of agent 1 is equal to $(2-p)$. Suppose next that agent 1 report strategically $0$ for $o_1$. As the mechanism is non-wasteful, it gives $o_1$ to agent 2 and $o_2$ and $o_3$ to agent 1 with probability $1$. The (expected) utility of agent 1 is equal to $2$. This outcome is strictly greater than $(2-p)$ as $p\in (0,1)$. In the second case, the mechanism is deterministic and $p=0$. The mechanism gives $o_1$ to agent 2 and $o_2$ and $o_3$ to agent 1 with probability $1$. The (expected) utility of agent 2 is $1$. Suppose next that agent 2 report strategically $0$ for $o_1$. The mechanism gives $o_1$ to agent 1 and $o_2$ and $o_3$ to agent 2 with probability $1$. The (expected) utility of agent 2 is $2$. This is a strict improvement. Analogously, for the third case when $p=1$.
\myqed
\end{myproof}

\subsection{Envy-Freeeness}

We next confirm that \emph{no} mechanism exists which is guaranteed to return envy-free allocations even in ex ante sense, supposing agents bid sincerely. The key idea behind this result is that a given agent may like a given bundle of items but not the individual items in the bundle. By comparison, with additive utilities, the \mylike\ mechanism for example is envy-free ex ante \cite{aleksandrov2015ijcai}. 
 
\begin{mytheorem}\label{thm:two}
No non-wasteful mechanism for online fair division is envy-free ex post or even ex ante, even with monotone utilities with 0/1 marginals.
\end{mytheorem} 

\begin{myproof}
Let us consider agents 1 and 2, items $o_1$ and $o_2$ arriving from $(o_1,o_2)$. Consider $u_1(\emptyset)=u_1(\lbrace o_1\rbrace)=u_1(\lbrace o_2\rbrace)=0,u_1(O)=1$ and $u_2(\emptyset)=0,u_2(\lbrace o_1\rbrace)=u_2(\lbrace o_2\rbrace)=1,u_2(O)=2$. We note that an envy-free (offline) allocation gives one item to each agent. However, an online and non-wasteful mechanism gives deterministically both items to agent 2. Hence, agent 1 envies agent 2.
\myqed
\end{myproof}  

Interestingly, with identical monotone utilities, a distribution of allocations that is envy-free in expectation can always be returned. For example, consider the non-wasteful mechanism that allocates the current item to an agent who makes a non-zero marginal bid for it and so far has been allocated items with the least declared utility. 

\begin{quote}
\myminlike: At round $j\in [m]$, given $\pi\in\Pi_{j-1}$, we let $\mbox{Like}=\lbrace i |v_{i}(\pi_i\cup\lbrace o_j\rbrace) > v_i(\pi_i)\rbrace$ and $\mbox{MinLike}=\lbrace i | i \in \mbox{Like}, v_{i}(\pi_i)=\min_{k\in\mbox{\scriptsize Like}} v_{k}(\pi_k)\rbrace$. The mechanism gives $o_j$ to some $i\in \mbox{MinLike}$ with probability $(1/|\mbox{MinLike}|)$ if $\mbox{MinLike}\neq\emptyset$ and, otherwise, to some $i\in [n]$ with probability $1/n$.
\end{quote}

\begin{mytheorem}\label{thm:three}
With identical monotone utilities, the non-wasteful \myminlike\ mechanism is envy-free ex ante.
\end{mytheorem} 

\begin{myproof}
The proof of the result hinges on any pair of agents getting a particular bundle of items with the same probability. Pick agents $i,k$. We show $\overline{u}_{ii}(\Pi_{j})\geq \overline{u}_{ik}(\Pi_{j})$ for $j\in [1,m]$. Let $\Delta \overline{u}_{ikj}=\overline{u}_{ii}(\Pi_{j})-\overline{u}_{ik}(\Pi_{j})$. We have $\Delta \overline{u}_{ikj}=\sum_{\pi} p(\pi)\cdot u_{i}(\pi_i)-\sum_{\pi} p(\pi)\cdot u_{i}(\pi_k)$ where $\pi\in\Pi_j$. We derive the below expression for $\Delta \overline{u}_{ikj}$.

\begin{center}
$\Delta \overline{u}_{ikj}=\displaystyle\sum_{A\subseteq O,B\subseteq O\setminus A}$
$\bigg(\displaystyle \sum_{\pi_i=A,\pi_k=B} p(\pi)\cdot u_i(A)+
\displaystyle\sum_{\pi_i=B,\pi_k=A} p(\pi)\cdot u_i(B)$
$-\displaystyle\sum_{\pi_i=A,\pi_k=B} p(\pi)\cdot u_i(B)-
\displaystyle\sum_{\pi_i=B,\pi_k=A} p(\pi)\cdot u_i(A)\bigg)$
\end{center}

Pick an allocation $\pi\in \Pi_j$. Let agent $i$ get $A\subseteq O$, agent $k\not=i$ get $B\subseteq O\setminus A$ and each other agent $h\not=i,k$ get $\pi_h$ in $\pi$. By the symmetry of the utilities, there is another allocation, say $\pi^{\prime}\in\Pi_j$, such that $i$ get $B$, $k$ get $A$ and $h$ get $\pi_h$. With \myminlike, $p(\pi^{\prime})=p(\pi)$. Moreover, with this mechanism, the number of returned allocations that give $A$ to $i$ and $B$ to $k$ is equal to the number of returned allocations that give $B$ to $i$ and $A$ to $k$. Therefore, we derive $\Delta \overline{u}_{ikj}=0$ for each $j\in [m]$.
\myqed
\end{myproof}

Further, we consider two common approximations of envy-freeness ex post: EF1 and EFX \cite{budish2011,caragiannis2016}. However, many other such approximations that are stronger than EF1 have been studied in the recent years, e.g.\ GMMS, PMMS, EFL \cite{amanatidis2018,aziz2018aaai,barman2018aaai}.

\begin{mydefinition}$($\emph{EF up to some item, EF1}$)$
A mechanism is \emph{EF1} if, for each $\pi\in \Pi_m$ with $p(\pi)>0$, for all $i,k$ with $\pi_k\not=\emptyset$, $\exists o\in \pi_k$ with $u_i(\pi_i)\geq u_i(\pi_k\setminus\lbrace o\rbrace)$. 
\end{mydefinition}

\begin{mydefinition}$($\emph{EF up to any item, EFX}$)$
A mechanism is \emph{EFX} if, for each $\pi\in \Pi_m$ with $p(\pi)>0$, for all $i,k, o\in \pi_k$ with $u_i(o)>0$, $u_i(\pi_i)\geq u_i(\pi_k\setminus\lbrace o\rbrace)$. 
\end{mydefinition}

Unfortunately, we cannot guarantee to only return allocations that are even envy-free up to some item. This holds under very strong restrictions on the preference domain. Consequently, there are \emph{no} EF1 (and, therefore, GMMS, PMMS or EFL) mechanisms for our setting in general. 

\begin{mytheorem}\label{thm:four}
No non-wasteful mechanism for online fair division is EF1, even with identical monotone utilities with 0/1 marginals.
\end{mytheorem}

\begin{myproof}
Consider agents 1 and 2, items $o_1$ to $o_4$ and ordering $(o_1,o_2,o_3,o_4)$. Let $B\subseteq O$. If $|B|=1$, let $u(B)=1$. If $|B|=2$ and $o_1\in B$, let $u(B)=1$. If $|B|=2$ and $o_1\not\in B$, let $u(B)=2$. If $|B|=3$ and $B=\lbrace o_2,o_3,o_4\rbrace$, let $u(B)=3$. If $|B|=3$ and $B\not=\lbrace o_2,o_3,o_4\rbrace$, let $u(B)=2$. Also, let $u(O)=3$. By these preferences, a non-wasteful mechanism gives $o_1$ to agent 1 and $o_2,o_3,o_4$ to agent 2, or $o_1$ to agent 2 and $o_2,o_3,o_4$ to agent 1. WLOG, let agent 1 get $o_1$ and agent 2 get $o_2,o_3,o_4$. The utilities of agents $1$ and $2$ in this allocation are \num{1} and \num{3} respectively. The allocation is not envy-free because agent 1 envies agent 2. Moreover, the envy of agent 1 remains even after the removal of any single item from the bundle of agent 2. Consequently, the allocation is not EF1. However, we note that an EF1 (offline) allocation gives two items to each agent.
\myqed
\end{myproof}

By Theorem~\ref{thm:four}, the \myminlike\ mechanism is not EF1. The result in Theorem~\ref{thm:four} also contrasts with the offline setting where, with general monotone utilities, an EF1 allocation, bounding the envy from above by the maximum marginal utility of any agent for any item, can always be achieved \cite{lipton2004,plaut2018}.

\subsection{Pareto Efficiency}

We lastly consider Pareto efficiency supposing agents bid sincerely. In the offline setting with general monotone utilities, Pareto efficiency is guaranteed \cite{chevaleyre2008,nguyen2014}. In our setting, we show that there is \emph{no} mechanism that is Pareto efficient, even just ex ante. 

\begin{mytheorem}\label{thm:five}
No non-wasteful mechanism for online fair division is Pareto efficient ex post or even ex ante, even with identical monotone utilities. 
\end{mytheorem}

\begin{myproof}
Consider agents 1 and 2, items $o_1$ to $o_4$ and ordering $(o_1,o_2,o_3,o_4)$. The utilities are identical for each $B\subseteq O$. If $|B|=1$, let $u(B)$ be $2$ if $B=\lbrace o_3\rbrace$ or $B=\lbrace o_4\rbrace$, and $1$ otherwise. If $|B|=2$, let $u(B)$ be $1$ if $B=\lbrace o_1,o_2\rbrace$ and $2$ otherwise. If $|B|=3$, let $u(B)$ be $3$ if $B=\lbrace o_1,o_2,o_4\rbrace$ and $2$ otherwise. Also, let $u(\emptyset)=0$ and $u(O)=3$. Further, consider below all possible allocations.

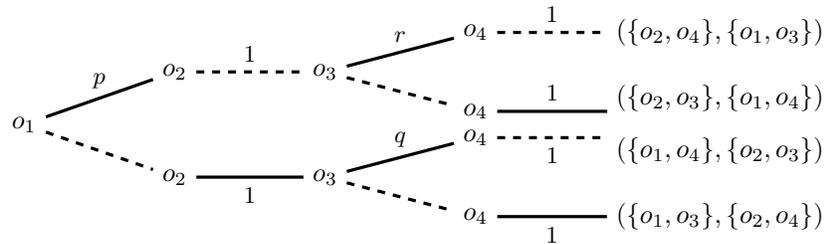
\begin{center}
\begin{tikzpicture}[xscale=1,yscale=0.7]
 
		\node at (0,1.75) (A) {$o_1$};
		
		\node at (2,2.750) (B) {$o_2$};
     	\node at (2,0.750) (C) {$o_2$};
     	 	
		\node at (4,2.750) (D) {$o_3$};
     	\node at (4,0.750) (E) {$o_3$};
     	
     	\node at (6,3.50) (F) {$o_4$};
     	\node at (6,2.00) (G)  {$o_4$};
    		\node at (6,1.50) (H)  {$o_4$};
       	\node at (6,0.00) (I) {$o_4$};
       	
       	\node at (9.25,3.50) (J) {$(\lbrace o_2,o_4\rbrace,\lbrace o_1,o_3\rbrace)$};
     	\node at (7.875,2.00) (K)  {};
    		\node at (7.875,1.50) (L)  {};
       	\node at (9.25,0.00) (M) {$(\lbrace o_1,o_3\rbrace,\lbrace o_2,o_4\rbrace)$};
     	\node at (9.25,2.25) (N)  {$(\lbrace o_2,o_3\rbrace,\lbrace o_1,o_4\rbrace)$};
    		\node at (9.25,1.25) (O)  {$(\lbrace o_1,o_4\rbrace,\lbrace o_2,o_3\rbrace)$};
       	
    	    \path[every node/.style={font=\sffamily\small}]
    (A) edge [black,very thick] node [black,above] {$p$} (B)
    (A) edge [black,very thick,dashed] node [black,below] {} (C)
    (B) edge [black,very thick,dashed] node [black,above] {$1$} (D)
    (C) edge [black,very thick] node [black,below] {$1$} (E)
    (D) edge [black,very thick] node [black,above] {$r$} (F)
    (D) edge [black,very thick,dashed] node [black,above] {} (G)
    (E) edge [black,very thick] node [black,above] {$q$} (H)
    (E) edge [black,very thick,dashed] node [black,below] {} (I)
    (F) edge [black,very thick,dashed] node [black,above] {$1$} (J)
    (G) edge [black,very thick] node [black,above] {$1$} (K)
    (H) edge [black,very thick,dashed] node [black,below] {$1$} (L)
    (I) edge [black,very thick] node [black,below] {$1$} (M);

\end{tikzpicture}
\emph{\newline Key: agent $1$-dashed line, agent $2$-solid line}
\end{center}

Each mechanism induces some probabilities $p,r,q\in [0,1]$. Such a mechanism allocates deterministically $o_2$ and $o_4$ to agents. For example, suppose that agent 2 get $o_1$ with probability $p$. Then, agent 1 gets $o_2$ with probability \num{1}. Suppose that agent 2 gets $o_3$ with probability $r$. Then, agent 1 gets $o_4$ with probability \num{1}. Each agent receives utility of $2$ in each of the four allocations. Hence, the agents' (expected) utilities are both equal to $2$. These allocations are Pareto dominated by $(\lbrace o_1,o_2,o_4\rbrace,\lbrace o_3\rbrace)$ in which agents 1 and 2 get utilities $3$ and $2$ respectively. The result follows.
\myqed
\end{myproof}

\section{Non-zero Marginal Utilities}

We continue with non-zero marginal utilities. Interestingly, we can achieve most axiomatic properties in this domain. Suppose we are interested in strategy-proofness, Pareto efficiency ex post and ex ante. Consider a simple mechanism that gives deterministically each next item to some fixed agent, say $i\in [n]$. Potentially, agent $i$ may wish to manipule the outcome. However, they then could only receive less items and, therefore, strictly less utility. Consequently, this mechanism is strategy-proof and, for the same reason, it is Pareto efficient even ex ante. Suppose we wish to achieve strategy-proofness, Pareto efficiency ex post and envy-freeness ex ante. Consider a mechanism that picks an agent, say $i\in [n]$, uniformly at random with probability $\frac{1}{n}$ and then gives deterministically each next item to $i$. This mechanism is strategy-proof and Pareto efficient ex post for the reasons that we mentioned above. It is further envy-free ex ante as it returns a distribution of $n$ allocations (say $\pi^i$ for $i\in [n]$ that occurs with probability $\frac{1}{n}$ and, WLOG, gives all items to agent $i$) in which the expected utility of an agent for their own allocation and the allocation of another agent is the same.

Unfortunately, both of the above mechanisms are unappealing because they give all items to some agent. Therefore, they are not EFX or even just EF1. In our online and monotone setting, there are \emph{no} mechanisms that are EF1 even when the utilities are positive and additive, a special case of non-zero marginal utilities. 

\begin{mytheorem}\label{thm:six}
No mechanism for online fair division is EF1, even with positive additive utilities. 
\end{mytheorem}

\begin{myproof}
Let us consider agents 1 and 2, items $o_1$ to $o_3$ and ordering $(o_1,o_2,o_3)$. Further, consider a mechanism and suppose that it is EF1. We consider two cases. In the first one, we assume that it gives $o_1$ to agent 1 with positive probability. Then, the utilities of agents for items are given in the below table.

\begin{center}
\begin{tabular}{|c|c|c|c|} \hline
& $o_1$ & $o_2$ & $o_3$ \\ \hline
agent 1 & \num{50} & \num{100} & \num{100} \\
agent 2 & \num{100} & \num{50} & \num{100} \\ \hline
\end{tabular}
\end{center}

WLOG, we can assume that the mechanism allocates $o_1$ at the first round. As it is EF1, it gives $o_2$ to agent 2. Given this partial allocation, there are only two possible allocations of $o_3$, resulting in $(\lbrace o_1,o_3\rbrace,\lbrace o_2\rbrace)$ and $(\lbrace o_1\rbrace,\lbrace o_2,o_3\rbrace)$. It is easy to check that none of them is EF1. 

In the second case, we assume that the mechanism gives $o_1$ to agent 2 with probability 1. Then, we consider different utilities of the agents for items $o_2$ and $o_3$. These are given in the below table.

\begin{center}
\begin{tabular}{|c|c|c|c|} \hline
& $o_1$ & $o_2$ & $o_3$ \\ \hline
agent 1 & \num{50} & \num{40} & \num{410} \\
agent 2 & \num{100} & \num{200} & \num{200} \\ \hline
\end{tabular}
\end{center}

The mechanism gives $o_1$ to agent 2. As it is EF1, it would then give $o_2$ to agent 1. Given this partial allocation, the only two possible allocations after the third round are $(\lbrace o_2\rbrace,\lbrace o_1,o_3\rbrace)$ and $(\lbrace o_2,o_3\rbrace,\lbrace o_1\rbrace)$. It is easy to check that neither of them is EF1.\myqed
\end{myproof}

In contrast, a simple \emph{round-robin} procedure returns an EF1 allocation in the offline setting with general additive utilities \cite{caragiannis2016}. There is some more hope for restricted preference domains on which to achieve EF1. For example, EF1 can be guaranteed in the special case of identical monotone utilities with non-zero marginals.

\begin{mytheorem}\label{thm:seven}
With identical monotone utilities with non-zero marginals, the non-waste\-ful \myminlike\ mechanism is EF1.
\end{mytheorem}

\begin{myproof}
We use induction on $j\in [m]$. In the base case, the allocation of $o_1$ is trivially EF1. In the step case, the induction hypothesis requires that $\pi\in\Pi_{j-1}$ with $p(\pi)>0$ is EF1. Let $1\in \mbox{MinLike}$ and the mechanism allocate $o_j$ to agent 1 given $\pi$. Consider $\pi^{\prime}=(\pi^{\prime}_1,\ldots,\pi^{\prime}_n)$ where $\pi^{\prime}_1=\pi_1\cup\lbrace o_j\rbrace$ and $\pi^{\prime}_i=\pi_i$ for each $i\not=1$. We next show that $\pi^{\prime}$ is EF1. We note that the set $\mbox{Like}=[n]$ as the agents' marginal utilities are non-zero.

\emph{Case 1}: Suppose $i\neq 1$ and $k \neq 1$. We have $u_i(\pi^{\prime}_i)=u_i(\pi_i)$ and $u_k(\pi^{\prime}_k)=u_k(\pi_k)$ as $\pi^{\prime}_i=\pi_i$ and $\pi^{\prime}_k=\pi_k$. By the hypothesis, we have $u_i(\pi_i)\geq u_i(\pi_k\setminus\lbrace o\rbrace)$ for some $o\in \pi_k\not=\emptyset$. Hence, $u_i(\pi^{\prime}_i)\geq u_i(\pi^{\prime}_k\setminus\lbrace o\rbrace)$ holds. Or, agent $i$ is EF1 of agent $k$ in $\pi^{\prime}$.

\emph{Case 2}: Suppose $i \neq 1$ and $k=1\in \mbox{MinLike}$. We have $u_i(\pi^{\prime}_i)=u_i(\pi_i)$ as $\pi^{\prime}_i=\pi_i$. By the mechanism, we have $u_i(\pi_i)\geq u_1(\pi_1)$. As the utilities are identical, we have $u_1(\pi_1)=u_i(\pi_1)$. Hence, $u_i(\pi_i)\geq u_i(\pi_1)$, or agent $i$ is envy-free of agent 1 in $\pi$. We derive $u_i(\pi^{\prime}_i)\geq u_i(\pi_1)=u_i(\pi^{\prime}_1\setminus\lbrace o_j\rbrace)$ as $\pi^{\prime}_1=\pi_1\cup\lbrace o_j\rbrace$. Hence, agent $i$ is EF1 of agent 1 in $\pi^{\prime}$.

\emph{Case 3}: Suppose that $i=1\in \mbox{MinLike}$ and $k \neq 1$. We have $u_1(\pi^{\prime}_1)>u_1(\pi_1)$ as $\pi^{\prime}_1=\pi_1\cup\lbrace o_j\rbrace$ and the utilities are with non-zero marginals. By the hypothesis, $u_1(\pi_1)\geq u_1(\pi_k\setminus\lbrace o\rbrace)$ for some $o\in\pi_k\not=\emptyset$. Hence, $u_1(\pi^{\prime}_1)>u_1(\pi_k\setminus\lbrace o\rbrace)=u_1(\pi^{\prime}_k\setminus\lbrace o\rbrace)$ as $\pi^{\prime}_k=\pi_k$. Therefore, agent 1 is EF1 of agent $k$ in $\pi^{\prime}$.
\myqed
\end{myproof}

By Theorem~\ref{thm:three}, the \myminlike\ mechanism is envy-free ex ante with identical monotone utilities with non-zero marginals. However, it is not strategy-proof. In fact, \emph{no} other EF1 mechanism satisfies this property.

\begin{mytheorem}\label{thm:eight}
No mechanism for online fair division is EF1 and strategy-proof, even with identical additive utilities.
\end{mytheorem}

\begin{myproof}
Let us consider two agents, items $o_1$ and $o_2$ arriving in $(o_1,o_2)$. Further, let both agents value $o_1$ with \num{1} and $o_2$ with \num{2}. We consider two cases. In the first one, suppose that the mechanism is randomized and allocates $o_1$ to agent 1 with probability $p\in (0,1)$ supposing agents 1 and 2 declare their sincere utilities for $o_1$ and $o_2$. Suppose it gives $o_1$ to agent 1. As the mechanism is EF1, it must give $o_2$ to agent 2 with probability of \num{1}. Suppose it gives $o_1$ to agent 2. As the mechanism is EF1, it must give $o_2$ to agent 1 with probability of \num{1}. Hence, agent 1 receives expected utility $(2-p)$. If agent 1 report strategically 0 for $o_1$, then the mechanism gives $o_1$ to agent 2 and $o_2$ to agent 1 with probability \num{1}. The expected utility of agent 1 is now \num{2} which is strictly higher than $(2-p)$ as $p>0$. Hence, the mechanism is not strategy-proof. In the second case, suppose that the mechanism is deterministic and allocates $o_1$ to agent 1 with probability \num{1}. Therefore, as it is EF1, it then allocates $o_2$ to agent 2 again with probability \num{1}. The utility of agent 1 in this returned allocation is \num{1}. If agent 1 report strategically 0 for $o_1$, then the mechanism swaps the items of the agents. The utility of agent 1 is now \num{2}. This is a strict improvement. We reached contradictions in both cases.\myqed
\end{myproof}

By Theorems~\ref{thm:three} and~\ref{thm:eight}, we conclude that the \myminlike\ mechanism returns an EF1 and envy-free ex ante allocation with identical additive utilities. In this case, the agents' utilities in each allocation is equal to the total sum of an agent's utilities for the items. For this reason, the mechanism is also Pareto efficient ex post and ex ante in this case. Unfortunately, this no longer holds whenever the utilities are monotone.

\begin{mytheorem}\label{thm:nine}
No mechanism for online fair division is EF1 and Pareto efficient ex post or even ex ante, even with identical monotone utilities with non-zero marginals.
\end{mytheorem}

\begin{myproof}
Let us consider two agents, items $o_1$ to $o_3$ arriving in $(o_1,o_2,o_3)$. The utilities are given in the below table.

\begin{center}
\begin{tabular}{|c|c|c|c|c|c|c|c|} \hline
& $o_1$ & $o_2$ & $o_3$ & $\lbrace o_1,o_2\rbrace$ & $\lbrace o_1,o_3\rbrace$ & $\lbrace o_2,o_3\rbrace$ & $O$\\ \hline
agent 1 & \num{1} & \num{2} & \num{3} & \num{4} & \num{4} & \num{4} & \num{5} \\
agent 2 & \num{1} & \num{2} & \num{3} & \num{4} & \num{4} & \num{4} & \num{5} \\ \hline
\end{tabular}
\end{center}

Let us consider a mechanism that gives item $o_1$ to agent 1 with probability $p\in [0,1]$. Suppose agent 1 receives item $o_1$. As the mechanism is EF1, it then gives deterministically item $o_2$ to agent 2 and item $o_3$ to agent 1. Hence, the allocation $\pi_1=(\lbrace o_1,o_3\rbrace,\lbrace o_2\rbrace)$ is returned with probability $p$. Suppose agent 2 receives item $o_1$. By the symmetry of the preferences, we conclude that the allocation $\pi_2=(\lbrace o_2\rbrace,\lbrace o_1,o_3\rbrace)$ is returned with probability $(1-p)$. We observe that $\pi_1$ is Pareto dominated by $\pi_3=(\lbrace o_1,o_2\rbrace,\lbrace o_3\rbrace)$ and $\pi_2$ is Pareto dominated by $\pi_4=(\lbrace o_3\rbrace,\lbrace o_1,o_2\rbrace)$. Hence, the mechanism is not Pareto efficient ex post. Further, with the mechanism, the expected utilities of agents 1 abd 2 are $(2+2\cdot p)$ and $(4-2\cdot p)$ respectively. For $p\geq [\frac{1}{2},1)$, the first of these outcomes is less than $4$ and the second one is at most $3$. For $p=1$, they are $4$ and $2$. These expected allocations are Pareto dominated by $\pi_3$ in which agent 1 receive utility $4$ and agent 2 receive utility $3$. For $p\in (0,\frac{1}{2})$, the first expected outcome is less than $3$ and the second one is less than $4$. For $p=0$, they are $2$ and $4$. These expected allocations are Pareto dominated by $\pi_4$ in which agent 1 receive utility $3$ and agent 2 receive utility $4$. Hence, the mechanism is not Pareto efficient ex ante.
\myqed
\end{myproof}

In the offline setting, an EF1 (even EFX) and Pareto efficient ex post (and, therefore, Pareto efficient ex ante) allocation can always be returned with identical monotone utilities whose marginals are non-zero \cite{plaut2018}. Further, by Theorem~\ref{thm:six}, we cannot even hope for mechanisms that satisfy the stronger concept of EFX with positive additive utilities. In fact, this holds even in the more special case of identical utilities. This contrasts with the offline setting \cite{barman2018}.

\begin{mytheorem}\label{thm:ten}
No non-wasteful mechanism for online fair division is EFX, even with identical additive utilities. 
\end{mytheorem}

\begin{myproof}
Consider agents 1 and 2, items $o_1$ to $o_3$ and $(o_1,o_2,o_3)$. For $i\in\lbrace 1,2,3\rbrace$, let each agent have utility $i$ for item $o_i$. We note that two EFX allocations exist: $(\lbrace o_1,o_2\rbrace,$ $\lbrace o_3\rbrace)$ and $(\lbrace o_3\rbrace,\lbrace o_1,o_2\rbrace)$. Consider a non-wasteful mechanism and suppose that it is EFX. Hence, it would give $o_1$ and $o_2$ to different agents because it is online and cannot predict that $o_3$ will also arrive. WLOG, let agent 1 get $o_1$ and agent 2 get $o_2$. Given this allocation, it is easy to see that any allocation of $o_3$ leads to a violation of EFX.
\myqed
\end{myproof}

\section{Extensions}

In this section, we consider several extensions of our work as a response to our impossibility results in the previous sections, that highlight the technical difficulty of our online and monotone setting.

\subsection{Online Strategy-Proofness}

In deciding if agents have any incentive to misreport preferences in an online setting, we may consider the past fixed but the future unknown. Indeed, we might not know what items will arrive next, or even if any more items will arrive. This leads to a \emph{new} and weaker form of \emph{online strategy-proofness}.

\begin{mydefinition} $($\emph{Online strategy-proofness, OSP}$)$
A mechanism is \emph{OSP} in a problem with $m$ items if, for each item $o_j\in O$, fixed information of $o_1$ to $o_{j-1}$ and no information of $o_{j+1}$ to $o_m$, no agent $i$ can strictly increase $\overline{u}_{i}(\Pi_j)$ by misreporting $u_i(\pi_i\cup\lbrace o_j\rbrace)-u_i(\pi_i)$ given any allocation $\pi\in\Pi_{j-1}$, supposing that agent $i$ bid sincerely their marginal utilities for $o_1$ to $o_{j-1}$ and each agent $k \neq i$ bid sincerely their marginal utilities for $o_1$ to $o_j$.
\end{mydefinition}

Interestingly, the \myminlike\ mechanism is online strategy-proof. The key idea is that the probability of an agent for each next item given an allocation of the past items is constant for each their positive marginal bid, supposing all other bids are fixed.

\begin{mytheorem}\label{thm:eleven}
The non-wasteful \myminlike\ mechanism is online strategy-proof.
\end{mytheorem}

\begin{myproof}
Consider a problem of $m$ items. Let us pick an arbitrary round $j\in [m]$, allocation $\pi\in\Pi_{j-1}$ and agent $i\in [n]$. We consider two cases. In the first one, $i\not\in\mbox{MinLike}$. Then, this agent cannot increase $\overline{u}_{i}(\Pi_j)$ by misreporting $u_i(\pi_i\cup\lbrace o_j\rbrace)-u_i(\pi_i)$ because, for any such misreported value, they remain outside $\mbox{MinLike}$. In the second case, $i\in\mbox{MinLike}$. Hence, they receive $o_j$ with probability $1/|\mbox{MinLike}|$ supposing they bid $u_i(\pi_i\cup\lbrace o_j\rbrace)-u_i(\pi_i)$ that is positive. In fact, this holds for any other positive marginal bid that they report for this item. However, this probability becomes \num{0} whenever they report zero marginal bid for the item. We conclude that $\overline{u}_{i}(\Pi_j)$ cannot increase.
\myqed
\end{myproof}

\subsection{Wasteful Mechanisms}

We say that a mechanism is \emph{wasteful} iff it is not non-wasteful. Clearly, \emph{no} wasteful mechanism is Pareto efficient ex post or even ex ante simply because one can improve the outcome of the mechanism by taking an item that is allocated to an agent who report a zero marginal bid for it and giving it to some other agent who make a positive marginal bid for the item. We, therefore, focus on envy-freeness and strategy-proofness. Let us consider the \emph{uniform mechanism} that gives each next item to an agent with probability $\frac{1}{n}$ given any allocation of past items. This mechanism is strategy-proof and envy-free ex ante because no agent can increase their own outcome and each agent receives the same probability for a given bundle of items. By Theorem~\ref{thm:six}, no wasteful mechanism is EF1 in general. However, we can bound the envy ex post with identical monotone utilities. For example, consider the wasteful version of the \myminlike\ mechanism, i.e.\ the \myminutil\ mechanism. This one is EF1 in this domain. 

\begin{quote}
\myminutil: At round $j\in [m]$, given $\pi\in\Pi_{j-1}$, we let $\mbox{MinUtil}=\lbrace i | i \in [n], v_{i}(\pi_i)=\min_{k\in [n]} v_{k}(\pi_k)\rbrace$. The mechanism gives $o_j$ to some $i\in \mbox{MinUtil}$ with probability $(1/|\mbox{MinUtil}|)$.
\end{quote}

\begin{mytheorem}\label{thm:twelve}
With identical monotone utilities, the wasteful \myminutil\ mechanism is EF1.
\end{mytheorem}

\begin{myproof}
We can use induction on $j\in [m]$ as in the proof of Theorem~\ref{thm:seven}. In the base case, the allocation of $o_1$ is trivially EF1. In the step case, consider $\pi^{\prime}=(\pi^{\prime}_1,\ldots,\pi^{\prime}_n)$ where $\pi^{\prime}_1=\pi_1\cup\lbrace o_j\rbrace$ and $\pi^{\prime}_i=\pi_i$ for each $i\not=1$, supposing that $\pi\in\Pi_{j-1}$ with $p(\pi)>0$ is EF1. We next show that $\pi^{\prime}$ is EF1. Suppose $i\neq 1$ and $k \neq 1$. This follows by Case 1 in Theorem~\ref{thm:seven}. Suppose $i \neq 1$ and $k=1\in \mbox{MinUtil}$. This follows by Case 2 in Theorem~\ref{thm:seven}. Suppose that $i=1\in \mbox{MinUtil}$ and $k \neq 1$. We have $u_1(\pi^{\prime}_1)\geq u_1(\pi_1)$ as $\pi^{\prime}_1=\pi_1\cup\lbrace o_j\rbrace$. As $\pi$ is EF1, $u_1(\pi_1)\geq u_1(\pi_k\setminus\lbrace o\rbrace)$ for some $o\in\pi_k\not=\emptyset$. Hence, $u_1(\pi^{\prime}_1)\geq u_1(\pi_k\setminus\lbrace o\rbrace)=u_1(\pi^{\prime}_k\setminus\lbrace o\rbrace)$ as $\pi^{\prime}_k=\pi_k$. Therefore, agent 1 is EF1 of agent $k$ in $\pi^{\prime}$. We conclude that $\pi^{\prime}$ is EF1.
\myqed
\end{myproof}

It is easy to see that the \myminutil\ mechanism is online strategy-proof with general utilities and envy-free ex ante with identical utilities. However, by Theorems~\ref{thm:eight}, ~\ref{thm:nine} and~\ref{thm:ten}, we conclude that \emph{no} wasteful mechanism, including \myminutil, is EF1 and strategy-proof or EF1 and Pareto efficient, or just EFX. 

As strategy-proofness is possible (e.g.\ the uniform mechanism), we might wish to achieve even a stronger form of strategic robustness. For example, group strategy-proofness captures the ability of groups of agents to manipulate mechanisms in their joint favor \cite{aleksandrov2017ijcai}. 

\begin{mydefinition} $($\emph{Group strategy-proofness, GSP}$)$
A mechanism is \emph{GSP} in a problem with $m$ items if, with complete information of $o_1$ to $o_m$, no group of agents $G$ can strictly increase $\sum_{i\in G}\overline{u}_{i}(\Pi_m)$ by misreporting their marginal bids for one or more item $o_j$ and allocation $\pi\in\Pi_{j-1}$, supposing that every agent $k\not\in G$ bid sincerely their marginal utilities for items $o_1$ to $o_m$.
\end{mydefinition}

Surprisingly, the (wasteful) uniform mechanism is group strategy-proof in general as the outcome of a given group can only decrease supposing some agents from the group bid strategically marginal zeros for some items, and cannot increase if some of these agents bid strategically any combination of positive bids for some of these items. By comparison, \emph{no} non-wasteful mechanism is group strategy-proof even with two agents who cooperate and have different positive utilities for one item \cite{aleksandrov2017ijcai}. However, it remains an interesting \emph{open} question if group strategy-proofness is achievable with a non-wasteful mechanism in the case of identical monotone utilities with non-zero marginals, or identical additive utilities. Nevertheless, by Theorem~\ref{thm:eight}, such a mechanism cannot be EF1.

\section{Conclusions}\label{sec:con}

We consider a model for online fair division in which agents have monotone utilities for bundles of items. We studied common axiomatic properties of mechanisms for this model such as strategy-proofness, envy-freeness and Pareto efficiency. We analysed these properties for several utility domains, e.g.\ general marginal utilities, non-zero marginal utilities, identical utilities, etc. For non-wasteful mechanisms, most properties cannot be guaranteed. For wasteful mechanisms, most properties can be guaranteed in isolation. However, we also proved some impossibility results for combinations of axiomatic properties. We summarize our results in Table~\ref{tab:one}. 

\begin{table}\label{tab:one}
\caption{Key: \myimp\ - impossibility, $\checkmark$ - possibility, $+$ - discussion after, $-$ - discussion before.}
\resizebox{0.975\textwidth}{!}{
\begin{tabular}{c|C|C|C|C|C|C|C|C|}

\cline{2-9}
& \multicolumn{4}{c|}{non-wasteful mechanisms} & \multicolumn{4}{c|}{wasteful mechanisms} \\\cline{2-9}
 & \multicolumn{2}{c|}{ general utilities} & \multicolumn{2}{c|}{ identical utilities} & \multicolumn{2}{c|}{ general utilities} & \multicolumn{2}{c|}{ identical utilities}  \\\cline{2-9}
\multirow{2}{*}{property}  & possibly \num{0} marginals & non-zero marginals & possibly \num{0} marginals & non-zero marginals & possibly \num{0} marginals & non-zero marginals & possibly \num{0} marginals & non-zero marginals \\ \hline

 \multicolumn{1}{|c|}{OSP} & $\checkmark$ [T11] &$\checkmark$ [T11]& $\checkmark$ [T11]& $\checkmark$ [T11] & $\checkmark$ [T12]$^+$ & $\checkmark$ [T12]$^+$ & $\checkmark$ [T12]$^+$ & $\checkmark$ [T12]$^+$ \\ \hline
 \multicolumn{1}{|c|}{SP} & \myimp\ [T1] & $\checkmark$ [T6]$^-$ & \myimp\ [T1] & $\checkmark$ [T6]$^-$ & $\checkmark$ [T12]$^-$ & $\checkmark$ [T12]$^-$ & $\checkmark$ [T12]$^-$ & $\checkmark$ [T12]$^-$  \\ \hline
     \multicolumn{1}{|c|}{GSP} & \myimp\ [T1] & \myimp\ [T12]$^+$ & \myimp\ [T1] & open & $\checkmark$ [T12]$^+$ & $\checkmark$ [T12]$^+$ & $\checkmark$ [T12]$^+$ & $\checkmark$ [T12]$^+$ \\ \hline\hline

 \multicolumn{1}{|c|}{EF1} & \myimp\ [T4] & \myimp\ [T6] & \myimp\ [T4] & $\checkmark$ [T7] & \myimp\ [T6] & \myimp\ [T6] & $\checkmark$ [T12] & $\checkmark$ [T12] \\ \hline
 \multicolumn{1}{|c|}{EFX} & \myimp\ [T10] & \myimp\ [T10] & \myimp\ [T10] & \myimp\ [T10] & \myimp\ [T10] & \myimp\ [T10] & \myimp\ [T10] & \myimp\ [T10] \\ \hline
 \multicolumn{1}{|c|}{EFA } & \myimp\ [T2] & $\checkmark$ [T6]$^-$ & $\checkmark$ [T3] & $\checkmark$ [T3]  & $\checkmark$ [T12]$^-$ & $\checkmark$ [T12]$^-$ & $\checkmark$ [T12]$^-$ & $\checkmark$ [T12]$^-$  \\ \hline\hline

 \multicolumn{1}{|c|}{PEP+PEA} & \myimp\ [T5] & $\checkmark$ [T6]$^-$ & \myimp\ [T5]& $\checkmark$ [T6]$^-$ & \myimp\ [T12]$^-$ & \myimp\ [T12]$^-$ & \myimp\ [T12]$^-$ & \myimp\ [T12]$^-$ \\ \hline
\end{tabular}
}
\end{table}

We also proposed two new mechanisms - \myminlike\ and \myminutil\ - that satisfy a relaxed form of strategy-proofness in general as well as envy-freeness ex ante and ex post up to some item in two domains with identical utilities. We summarize these results in Table~\ref{tab:two}.

\begin{table}\label{tab:two}
\caption{Key: $\times$ - does not hold, $\checkmark$ - holds, $+$ - discussion after, $-$ - discussion before.}
\centering
\resizebox{0.825\textwidth}{!}{
\begin{tabular}{|c|c|c|c|c|c|c|c|}
\hline
mechanism & SP & OSP & EFA & EF1 & EFX & PEA & PEP \\\hline

& \multicolumn{7}{c|}{identical monotone utilities} \\\cline{1-8}
\myminlike & $\times$ [T1] & $\checkmark$ [T11] & $\checkmark$ [T3] & $\times$ [T4] & $\times$ [T10] & $\times$ [T5] & $\times$ [T5] \\\hline
\myminutil & $\times$ [T8] & $\checkmark$ [T12]$^+$ & $\checkmark$ [T12]$^+$ & $\checkmark$ [T12] & $\times$ [T10] & $\times$ [T12]$^-$ & $\times$ [T12]$^-$ \\\hline

& \multicolumn{7}{c|}{identical monotone utilities with non-zero marginals} \\\cline{1-8}
\myminlike & $\times$ [T8] & $\checkmark$ [T11] & $\checkmark$ [T3] & $\checkmark$ [T7] & $\times$ [T10] & $\times$ [T9] & $\times$ [T9] \\\hline

& \multicolumn{7}{c|}{identical additive utilities} \\\cline{1-8}
\myminlike & $\times$ [T8] & $\checkmark$ [T11] & $\checkmark$ [T3] & $\checkmark$ [T7] & $\times$ [T10] & $\checkmark$ [T9]$^-$ & $\checkmark$ [T9]$^-$ \\\hline
\end{tabular}
}
\end{table}

Finally, our results hold in offline fair division as well because online mechanisms can be applied to offline problems by picking up an order of the items. In future, we will consider other utility domains such as sub- and super-additive, or sub- and sup-modular utilities. We will also consider other relaxations of the considered properties and other (e.g.\ not marginal) mechanisms for our model.

\bibliographystyle{splncs}
\bibliography{online}

\end{document}